\documentclass[11pt]{iopart}
\usepackage{iopams}
\expandafter\let\csname equation*\endcsname\relax
\expandafter\let\csname endequation*\endcsname\relax
\usepackage{amsmath}
\usepackage{amsfonts}
\usepackage{amsthm}
\usepackage{dsfont}
\usepackage{cite}
\usepackage[pdftex]{graphicx}
\def\FF#1{\begin{eqnarray} #1 \end{eqnarray}}
\newcommand{\om}{\omega_0}
\newcommand{\OM}{\omega}

\newcommand{\N}{d}
\newcommand{\nn}{\nonumber}

\newcommand{\re}{{\rm e}}
\newcommand{\ri}{{\rm i}}
\def\tfrac12{\case{1}{2}}
\newcommand{\C}{\mathds{C}}
\newcommand{\R}{\mathds{R}}

\renewcommand{\Im}{\operatorname{Im}}
\renewcommand{\Re}{\operatorname{Re}}

\renewenvironment{pmatrix}{\left(\!\!\begin{array}{cc}}{\end{array}\!\!\right)}
\newenvironment{ppmatrix}{\left(\!\!\begin{array}{cccc}}{\end{array}\!\!\right)}
\begin{document}
\title{Classical and quantum dynamics in the (non-Hermitian) Swanson oscillator}
\author{Eva-Maria Graefe$^{1}$, Hans J\"urgen Korsch$^2$, Alexander Rush$^{1}$, and Roman Schubert$^{3}$}
\address{$^1$ Department of Mathematics, Imperial College London, London SW7 2AZ, UK }
\address{$^2$ FB Physik, TU Kaiserslautern, D--67653 Kaiserslautern, Germany}
\address{$^3$ Department of Mathematics, University of Bristol, Bristol, BS8 1TW, UK}
\begin{abstract}
The non-Hermitian quadratic oscillator known as Swanson oscillator is one
of the popular $PT$-symmetric model systems. Here
a full classical description of its dynamics is derived using recently
developed metriplectic flow equations, which combine the
classical symplectic flow for Hermitian systems with a dissipative metric flow
for the anti-Hermitian part. Closed form  expressions for 
the metric and phase-space trajectories are presented which are found to be periodic in time. Since the Hamiltonian is only quadratic the classical dynamics exactly describes the quantum dynamics of Gaussian wave packets.   
It is shown that the classical metric and trajectories as well as the
quantum wave functions can diverge in finite time even though the $PT$-symmetry is unbroken, i.e., the eigenvalues are purely real.
\end{abstract}

\section{Introduction}
In view of the increasing interest in non-Hermitian quantum systems \cite{Mois11book},
in particular $PT$-symmetric ones \cite{Bend05,Most10}, it is surprising
that only little is known hitherto about their classical limit,
despite of the fact that this limit is indispensable for a
description of Hermitian quantum dynamics. A first step towards 
classical non-Hermitian dynamics has been made by the authors 
in \cite{10nhclass,11nhcs}. The interesting complex structure
of the resulting metriplectic flow in phase space was revealed in \cite{12complex_coherent}.
Here we will analyse the non-Hermitian classical dynamics of one of the most
basic and most popular non-Hermitian model systems, a Hamiltonian quadratic in
position and momentum studied first by Ahmed and made popular by Swanson \cite{Ahme02,Swan04}. We will demonstrate that the
classical non-Hermitian description provides an important insight
into quantum dynamics and reveals features that are far from obvious. 
For example we will show that Gaussian wave packets 
diverge in finite time if the initial conditions are chosen in a critical region, even though 
the $PT$-symmetry is unbroken, and the system can be mapped to a Hermitian harmonic oscillator. 
This is related to the unboundedness of the operator mapping the $PT$-symmetric 
Hamiltonian to its Hermitian counterpart, that is typical for $PT$-symmetric systems on infinite dimensional Hilbert spaces \cite{Baga10,Bend12,Sieg12,Baga13,Baga13b,Bend13}. 

The paper is organised as follows: We introduce the quantum Hamiltonian in section \ref{sec_Ham} and give a brief review of its symmetry and spectral features. In section \ref{sec_class} we investigate the classical dynamics related to this system and provide analytic solutions for the phase space dynamics. As the Hamiltonian under consideration is quadratic in momentum and position, the classical dynamics also allows for an exact description of the quantum dynamics of Gaussian wave packets, which is presented in section \ref{sec_Qdyn}. We end with a brief conclusion. Details of the exact quantum time evolution with quadratic Hamiltonians and the semiclassical limit of phase space dynamics generated by non-Hermitian Hamiltonians are summarised in two appendices.
\section{The quantum Hamiltonian}
\label{sec_Ham}
We study a non-Hermitian $PT$-symmetric harmonic oscillator of the form 
\begin{equation}
\label{eqn_Ham}
\hat{\mathcal{H}} = \frac{\om}{2}\left(\hat{p}^2 + \hat{q}^2 \right)
 - \frac{i\delta}{2}(\hat{p}\hat{q} + \hat{q}\hat{p}),
\end{equation}
with $\delta,\om\in\mathds{R}$ and $\hbar=1$, which is $PT$ symmetric with the usual form of parity and time reversal operators
\begin{eqnarray}
P:\, \hat x\to-\hat x,\quad \hat p\to-\hat p\\
T:\, \hat x\to\hat x,\quad \hat p\to-\hat p,\quad \rmi\to-\rmi.
\end{eqnarray}
This Hamiltonian is unitarily equivalent to the Swanson Hamiltonian 
\begin{equation}
\hat{\mathcal H}_{swan}=\frac{a}{2}\hat p^2+\frac{b}{2}\hat q^2-\rmi\frac{\delta}{2}\left(\hat p\hat q+\hat q\hat p\right),
\end{equation}
for ${\rm sgn}(a)={\rm sgn}(b)$, via the transformation $\hat U\hat{\mathcal H}_{swan}\hat U^{-1}$ with 
\begin{equation}
\hat U={\rm e}^{\rmi\frac{1}{4}\ln\left(\frac{b}{a}\right)\left(\hat p\hat q+\hat q\hat p\right)},
\end{equation}
and with $\om={\rm sgn}(a)\sqrt{ab}$. In the following we shall confine the discussion to the case $\om>0$ for simplicity. 

From previous investigations of the Swanson oscillator it is known that the $PT$-symmetry is unbroken in the present case, the spectrum is discrete and harmonic with frequency $\omega=\sqrt{\om^2+\delta^2}$, and that the ground state is normalisable \cite{Ahme02,Swan04,Jone05,Musu07,Krej14}. As has been shown in \cite{Krej14} the Swanson oscillator is unitarily equivalent to a well studied complex deformed oscillator known as the Davies oscillator \cite{Davi99}, given by\begin{equation}
\label{eqn-Davies}
H_{\xi}:=\frac{1}{2}(\hat{p}^2+\xi^4\hat{q}^2),
\end{equation}
Specifically, the Hamiltonian (\ref{eqn_Ham}) is related to the Davies oscillator (\ref{eqn-Davies}) via
\begin{equation}
\hat{\mathcal{H}}=\frac{\omega_0}{\xi^{2}}\,\rme^{\frac{\rmi\pi}{8}(\hat p^2+\hat q^2)}H_{\xi}\rme^{-\frac{\rmi\pi}{8}(\hat p^2+\hat q^2)}
\end{equation}
where
\begin{equation}
\xi^4=\frac{\omega+\rmi \delta}{\omega-\rmi\delta}\,\, .
\end{equation}
This allows to directly transfer results for the Davies oscillator to the case studied here. While the spectrum is discrete and real, the system has a nontrivial pseudo spectrum, as has been discussed in detail in \cite{Krej14}. 

The Hamiltonian (\ref{eqn_Ham}) can also be mapped to a standard Hermitian harmonic oscillator
\begin{equation}
\label{eqn_HO}
\hat{H}_{herm}= \frac{\omega}{2}\left(\hat{p}^2 + \hat{q}^2\right),
\end{equation}
via the non-unitary similarity transformation $\hat{\mathcal{H}} = \hat\eta\hat{H}_{herm}\hat\eta^{-1}$, with 
\begin{equation}
\label{eqn_eta}
\hat \eta= \rme^{-\frac{\theta}{2}\left(\hat{p}^{2}-\hat{q}^{2}\right)},
\end{equation}
where $\theta$ is chosen such that,
\begin{equation}
\label{eqn_theta}
\tan\left(2\theta\right) =-\frac{\delta}{\om}.
\end{equation}
It then follows for the frequency of the Hermitian harmonic oscillator 
\begin{equation}
\label{eqn_om}
\omega=\sqrt{\om^2+\delta^2},
\end{equation}
and thus, the eigenvalues are given by
\begin{equation}
E_n=\omega (n+\tfrac12).\label{E-quant}
\end{equation}
Note that two operators related by an unbounded similarity transformation, such as $\hat\eta$ appearing here, are not necessarily isospectral. Nevertheless, in the present case we have isospectrality. Also the eigenstates of the Hamiltonian (\ref{eqn_Ham}) can be obtained by acting with $\hat\eta$ on the eigenstates of (\ref{eqn_HO}), i.e., the standard harmonic oscillator states, since these belong to the domain of $\hat \eta$. In particular, the ground state can easily be deduced as (see \ref{appendix_Gaussian_dyn} for details)
\begin{equation}
\psi(x)= \left(\frac{\om}{\pi(\omega-\delta)}\right)^{1/4} {\rm e}^{-\frac{\om}{2(\omega-\delta)} x^2}.
\end{equation}
Since $\omega\geq\delta$ by definition, $\frac{\om}{\omega-\delta}$ is positive, that is, the ground state is a Gaussian state for arbitrary parameters. 

There are an infinite number of alternative mappings to other isospectral Hermitian Hamiltonians (see \cite{Musu07,Jone05}), however, none of them is bounded, that is, the similarity transformation necessarily 
maps some states in $L^2(\mathds{R})$ out of $L^2(\mathds{R})$. The mathematical issue of the lack of an unbounded mapping to a Hermitian operator for PT-symmetric systems has been discussed extensively in the literature \cite{Baga10,Bend12,Sieg12,Baga13,Baga13b,Bend13,Most13,Krej14}. In what follows we shall show that this is related to an interesting dynamical phenomenon; in the classical dynamics the phase space trajectories periodically tend to infinity in finite time for particular initial conditions, and in the quantum dynamics, initially normalisable states momentarily leave $L^2(\mathds{R})$. 

\section{Classical metriplectic dynamics}
\label{sec_class}
Let us start by discussing the classical dynamics of our system, related to the 
dynamics of expectation values of the position and momentum operators in the Ehrenfest 
sense, as recently introduced in \cite{11nhcs}. The derivation is briefly summarised in \ref{app-semi}. Note that 
while the Hamiltonian is non-Hermitian, the operators $\hat p$ and $\hat q$ are Hermitian, and thus their expectation values are real. Thus, the classical counterpart of the quantum evolution is a non-Hamiltonian dynamics on a real phase space. The complex extension of Hamilton's equations leading to complex valued dynamical phase-space coordinates as discussed for example in \cite{Xavi96,00complex,Bend07b,Curt07,Gold08,Dey13}, however, is closely related to the dynamics on real phase space, and can be a useful tool for the calculation of the latter \cite{12complex_coherent} (see also \ref{appendix_Gaussian_dyn}). 

The classical counterpart of the Hamiltonian (\ref{eqn_Ham}) is given by 
\begin{eqnarray}
\label{nh-swan-4}
{\cal H}=H-\ri \Gamma
\quad \textrm{with}\quad
H(P,Q)=\frac{\om}{2}\left(P^2+Q^2\right)\quad,\quad
\Gamma(P,Q)=\delta PQ, 
\end{eqnarray}
where we use capital letters $P$, $Q$ to denote the real valued momentum and position variables. We reserve the notation $p$, $q$ for the complex phase-space variables discussed below. The classical equation of motion for the phase-space point $Z=(P,Q)$ is
\begin{eqnarray}
\label{nh-class-Z}
\dot Z=\Omega\nabla H -G^{-1}\nabla \Gamma
\end{eqnarray}
where $\nabla$ is the phase-space gradient, $\Omega$ is the standard symplectic unit matrix, 
\begin{equation}
\label{eqn_Omega}
\Omega=\begin{pmatrix}0 &-1\\1 &0\end{pmatrix},
\end{equation}
and the matrix $G$ is
the phase-space metric, compatible with the symplectic structure, $G\Omega G=\Omega$. In Hamiltonian dynamics the phase space is usually not thought of as being equipped with a metric. In the context of dissipative dynamics, however, the metric appears. This is not surprising from the point of view of complex geometry. It might seem natural to assume the metric to be given by the standard euclidean metric $G=I$ \cite{10nhclass}. As has been shown in \cite{11nhcs,12complex_coherent}, however, starting from the quantum dynamics in the semiclassical limit, $G$ is itself time dependent. The time evolution of the metric is governed by the dynamical equation 
\begin{eqnarray}
\label{nh-class-class-G}
\dot G=H''\Omega G-G\Omega H''+\Gamma''-G\Gamma''_\Omega G
\end{eqnarray}
where $H''$ and $\Gamma''$ are the matrices of second phase-space derivatives and
$\Gamma''_\Omega=\Omega^{\rm T}\Gamma''\Omega$. These are functions of the phase-space coordinate $Z$, that become constant for quadratic Hamiltonians. If the initial condition $G(0)$ is  a real symmetric matrix with determinant one (as a metric should be), these properties are automatically preserved in time for $G(t)$. In the general case the metric dynamics is coupled to the motion of the centre $Z$ via the functional dependence of $H=H(Z)$ and $\Gamma=\Gamma(Z)$. In the quadratic case discussed here, the dynamics of the metric is independent of  the phase-space dynamics, while the latter still depends on the metric. Note that the dynamical equations here also appear in the context of complexified geometric quantisation \cite{Burn13}.

In addition to the phase-space dynamics for non-Hermitian systems, the overall norm of the wave function, that is, the probability to find the quantum particle in the system, is time dependent. The dynamics of this survival probability $n(t)$, describing loss and gain dynamics, in the classical limit becomes
\begin{eqnarray}
\label{nh-class-class-n}
\dot n=-\big(2\Gamma +\tfrac12 {\rm tr}(\Gamma''_\Omega G)\big)\,n.
\end{eqnarray}
This can be integrated once the dynamics of $G$ and $Z$ have been found.

The phase-space dynamics is obtained from the solution of the coupled differential
equations (\ref{nh-class-Z}) for the phase-space variables  and  (\ref{nh-class-class-G}) for the 
metric, which in practice can be non-trivial, even for the case of an harmonic oscillator discussed here. In \cite{12complex_coherent} it has been shown that one can take advantage of 
an extension to a doubled four-dimensional phase space combined with projection techniques
relating complex valued solutions of the classical
time evolution to the desired real valued ones as follows. 
The metric $G(t)$ of the real valued two-dimensional phase space can be found as 
\begin{equation}
\label{eqn_Gdouble}
G(t)=\Phi(t)_*G(0):=
(\Phi_{pp}(t)G(0)+\Phi_{pq}(t))(\Phi_{qp}(t)G(0)+\Phi_{qq}(t))^{-1}
\end{equation}
from the solution of a dynamical equation in doubled phase space
\FF{\dot \Phi =\Omega_4K\Phi\quad {\textrm with} \quad \Phi(0)=I_4\,,\label{dotpHI}}
for the $4\times 4$ matrix
$\Phi(t)=\begin{pmatrix}\Phi_{pp}(t)&\Phi_{pq}(t)\\\Phi_{qp}(t)&\Phi_{qq}(t)\end{pmatrix}$,
where $I_4$ is the four-dimensional identity matrix and
\FF{\Omega_4=\begin{pmatrix}0 & -I\\I & 0\end{pmatrix}\ ,\quad
K=\begin{pmatrix}\Gamma''_{\Omega} & \Omega\,H''\\[2mm]
-H''\Omega &-\Gamma'' \end{pmatrix}\,.}
The real valued phase-space trajectory $Z(t)=(P(t),Q(t))^{\rm T}$ can then be found as 
\begin{equation}
\label{eqn_Zandz}
Z(t)=\Re\,z(t)-\Omega G(t)\,\Im\,z(t)\,,
\end{equation}
from the complex phase-space trajectory $z(t)=(p(t),q(t))^{\rm T}$ that solves the complex version of Hamilton's canonical equations
\begin{equation}
\dot z=\Omega\nabla {\mathcal H}.
\end{equation}
To every real phase space trajectory there corresponds a whole class of complex trajectories, that lead to the same projection (\ref{eqn_Zandz}). The set of equivalent complex phase space points at a given time forms a Lagrangian manifold \cite{12complex_coherent}.

In the present case we have 
\begin{eqnarray}
H''=\begin{pmatrix}\om & 0\\0 & \om\end{pmatrix}
\ ,\quad
\Gamma''=\begin{pmatrix}0 &\delta\\ \delta & 0\end{pmatrix}\,,
\end{eqnarray}
and thus
\FF{\Omega_4 K=
\begin{ppmatrix}
0 & -\om & 0 & \delta\\
\om & 0 & \delta & 0\\
0 & -\delta & 0 & -\om\\
-\delta & 0 & \om & 0 
\end{ppmatrix}.
}
Differentiating (\ref{dotpHI}) in time leads to 
\FF{\ddot \Phi=\Omega_4K''\dot \Phi=(\Omega_4 K'')^2\Phi=-\omega^2 \Phi\,,}
where $\omega=\sqrt{\om^2+\delta^2}$ is the frequency appearing in the quantum eigenvalues (\ref{E-quant}), 
and we have used that $(\Omega_4 K'')^2=-\omega^2I_4$. 
Thus, we find
\FF{\Phi(t)=\re^{\Omega_4 K'' t}=\cos \omega t\, I_4+\frac{\sin \omega t}{\omega}\,\Omega_4 K'',} 
that is,
\FF{\Phi_{pp}(t)&=\begin{pmatrix} 
\cos \omega t  & -\frac{\om}{\omega}\,\sin \omega t\\
\frac{\om}{\omega}\,\sin \omega t & \cos \omega t
\end{pmatrix}&=\Phi_{qq}(t)\,\\
\Phi_{pq}(t)&=\begin{pmatrix} 
0 &\quad \frac{\delta}{\omega}\,\sin \omega t\\
\frac{\delta}{\omega}\,\sin \omega t &\quad 0
\end{pmatrix}&=-\Phi_{qp}(t)\,.}
From these results the time dependent metric for arbitrary initial conditions can 
explicitly be expressed via equation (\ref{eqn_Gdouble}). 

The complex canonical equations for $p$ and $q$ are
\FF{\dot p &=&\ri\delta p-\om q \nn\\
\dot q&=&\om p-\ri\delta q,}
which can be directly solved to find the complex phase-space trajectories for the initial conditions $p(0)=p_0$ and $q(0)=q_0$ as
\begin{eqnarray}
p(t) &=&p_0\cos \OM t +\OM^{-1}\big(-\om q_0+\ri \delta p_0\big)\sin\OM t\nn\\
q(t) &=&q_0\cos \OM t +\OM^{-1}\big(\om p_0-\ri \delta q_0\big)\sin\OM t\,.
\label{comp-sol}
\end{eqnarray}
The time dependent metric and the real valued phase-space dynamics can be directly obtained via equations (\ref{eqn_Gdouble}) and (\ref{eqn_Zandz}), respectively. Once the solutions are known the dynamical equation for the survival probability 
\FF{\dot n=\delta\big(-2PQ-g_{pq}\big)n\,\label{dgl-norm}}
can be integrated as well. 

\begin{figure}[htb]
\begin{center}
\includegraphics[width=0.32\textwidth]{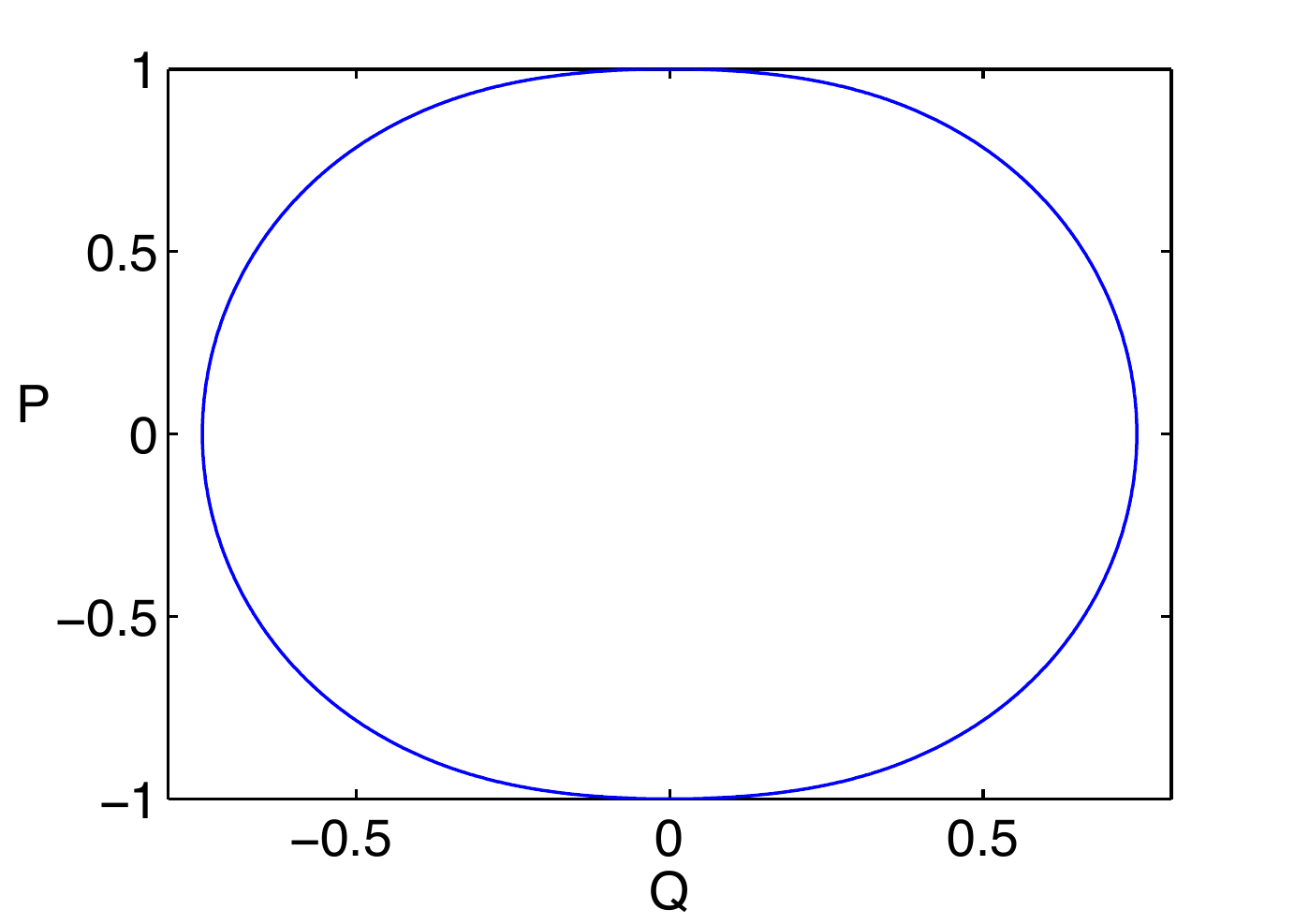}
\includegraphics[width=0.32\textwidth]{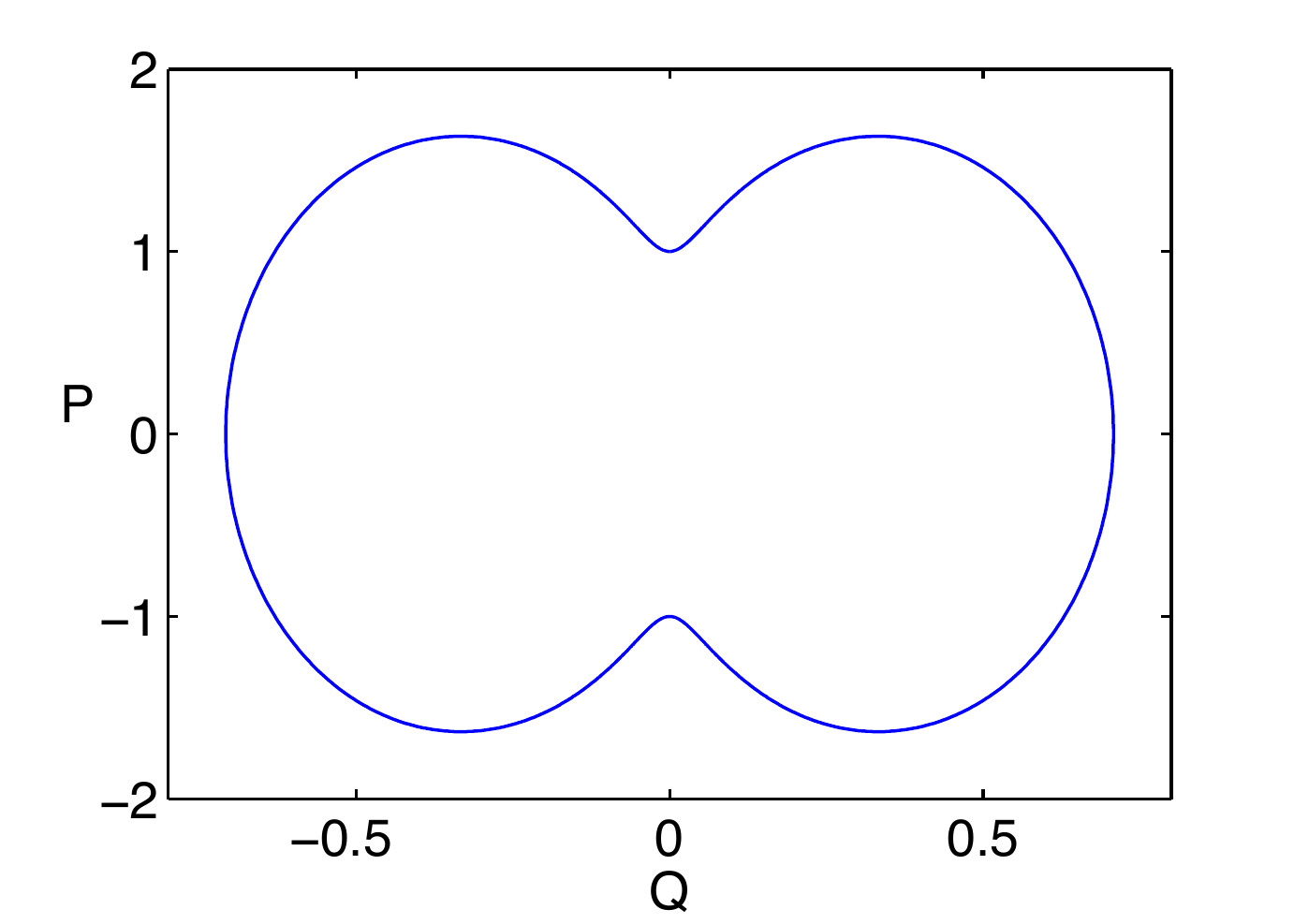}
\includegraphics[width=0.32\textwidth]{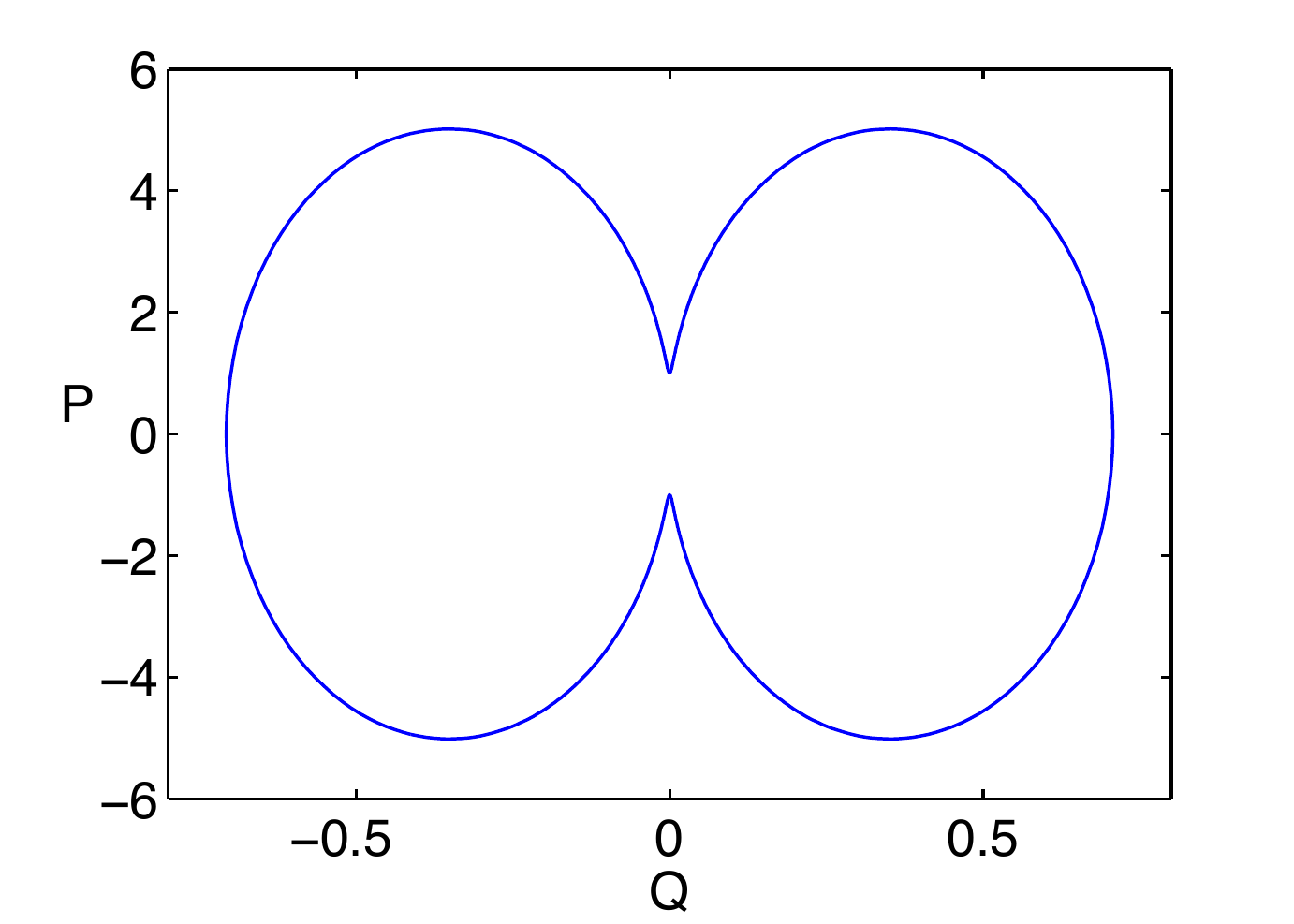}
\includegraphics[width=0.32\textwidth]{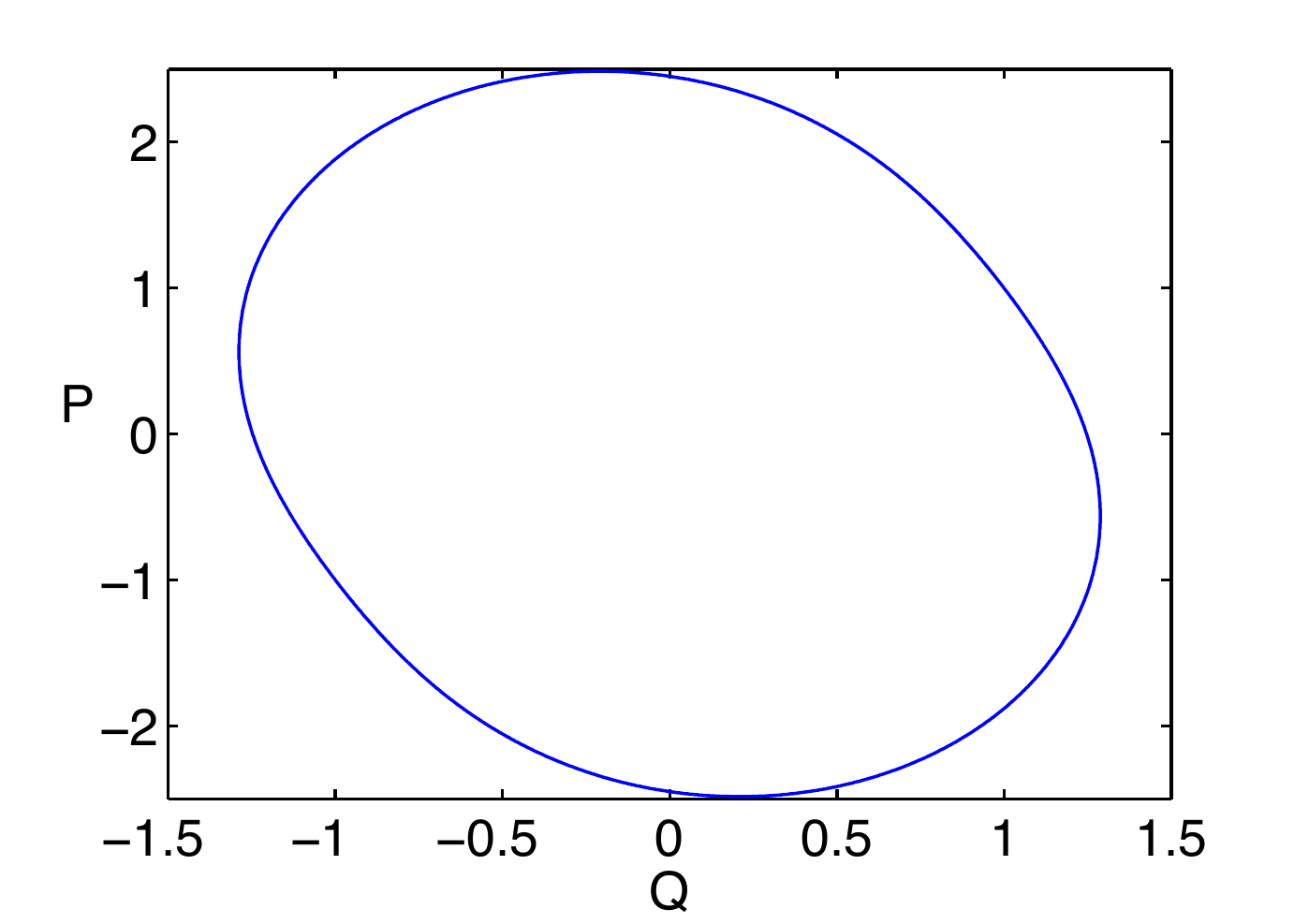}
\includegraphics[width=0.32\textwidth]{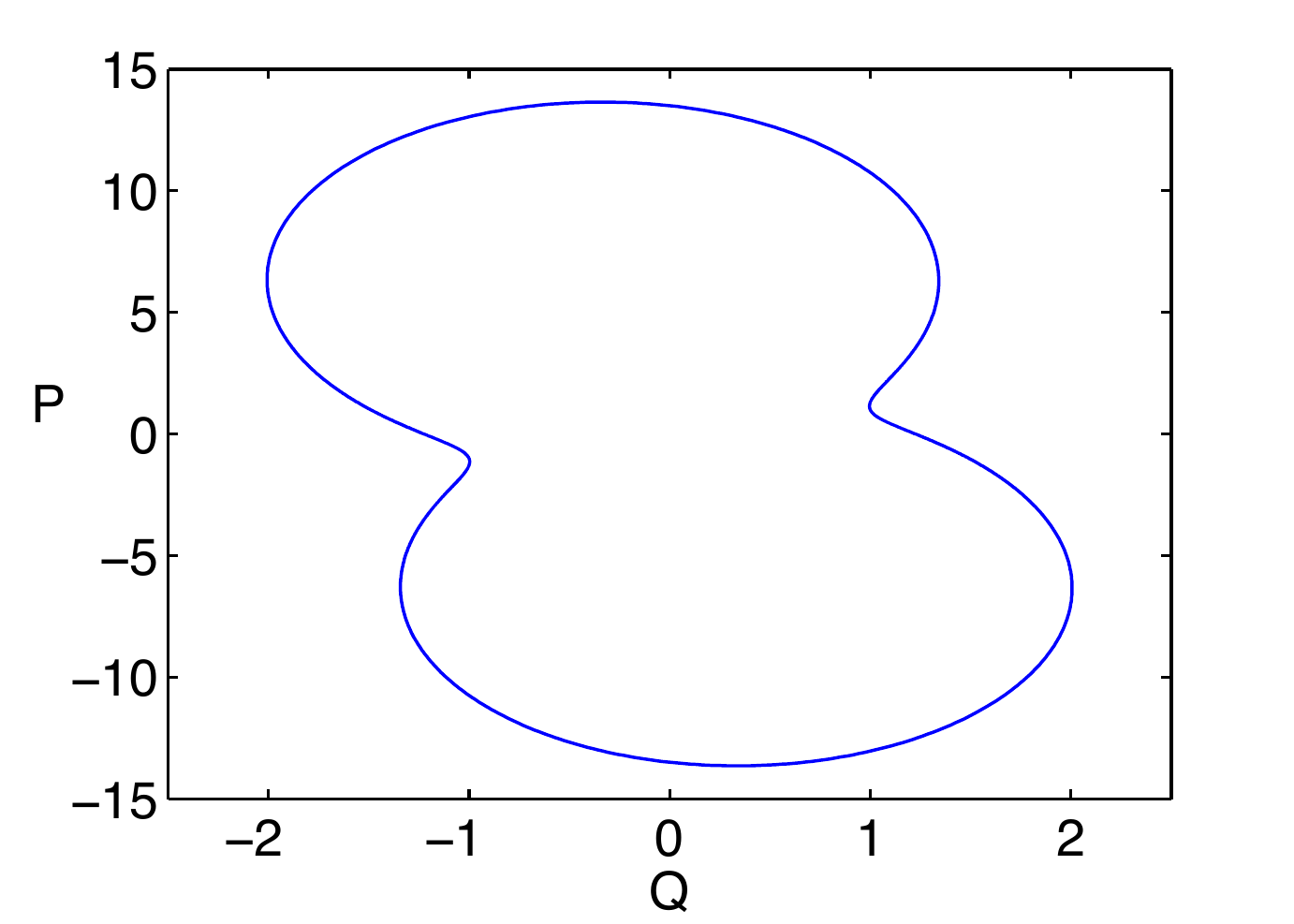}
\includegraphics[width=0.32\textwidth]{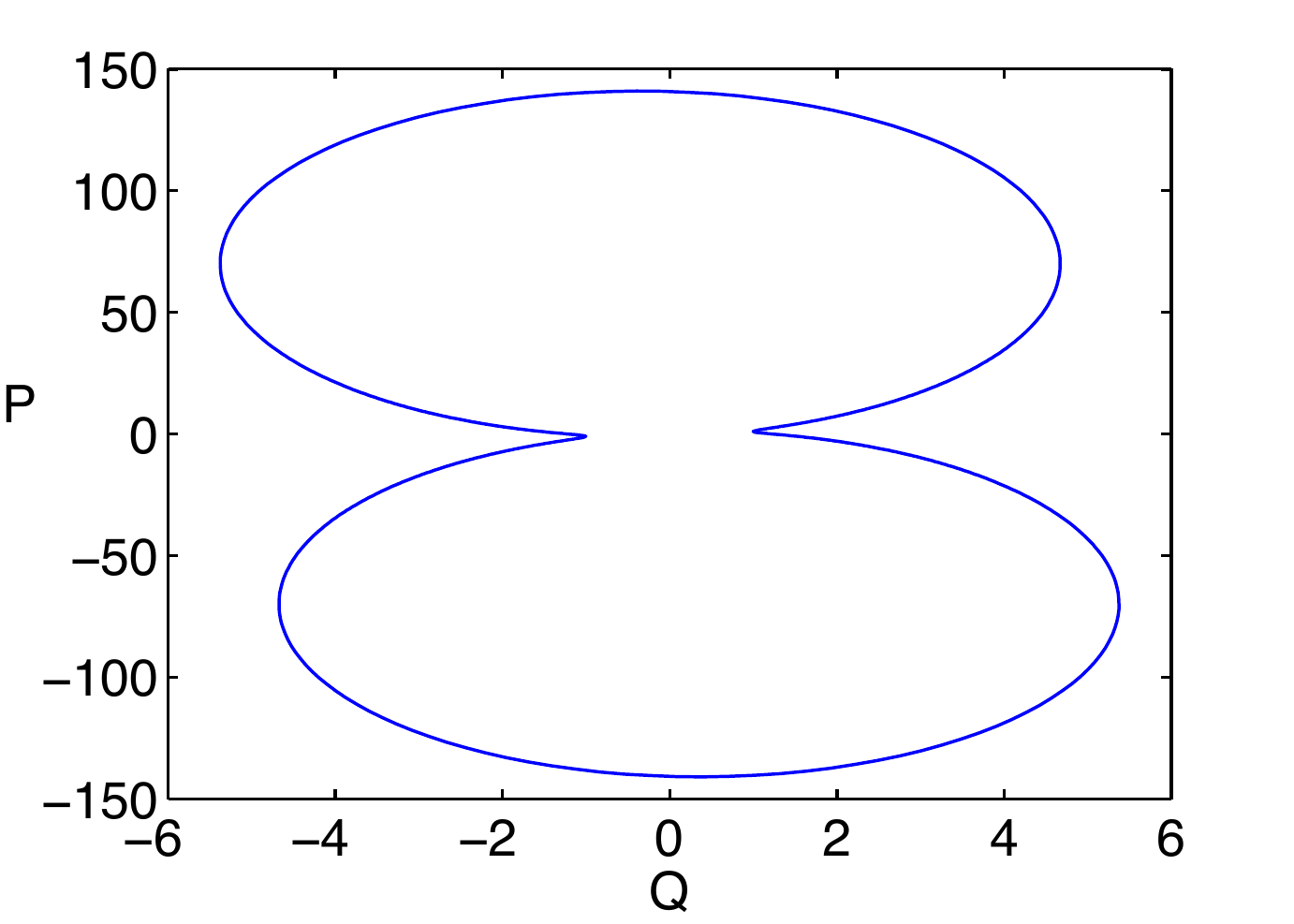}
\caption{Phase-space trajectories for different parameter values and initial conditions. The parameters are $\om=1$  $\delta=0.5,\,0.9,\,0.99$ (from left to right). The initial conditions for the top panel are $(P_0,Q_0)=(1,0)$ and for the bottom panel $(P_0,Q_0)=(1,1)$.}
\label{fig_PQdyn}
\end{center}
\end{figure}

The resulting dynamics is strictly periodic with frequency $\omega$, as it is expected due to the relation to a harmonic oscillator of that frequency. However, depending on the initial conditions and the values of the parameters $\om$ and $\delta$, divergences in the dynamical variables can occur. Let us illustrate this point by considering the initial condition $G(0)=I$. 
In this case equation 
(\ref{eqn_Gdouble}) simplifies to 
\FF{G(t)=(\Phi_{pp}(t)+\Phi_{pq}(t))
(\Phi_{qp}(t)+\Phi_{qq}(t))^{-1}\,.\label{Gmit_*}}
which yields
\FF{G(t)=\N(t)\!\left(\! \begin{pmatrix}1 &0\\ 0 & 1\end{pmatrix}
+\frac{\delta}{\omega^2}\begin{pmatrix}-\om(1-\cos 2\omega t) &\omega\sin 2\omega t\\ 
\omega\sin 2\omega t &  \om(1-\cos 2\omega t)\end{pmatrix}\!\right),\label{G-sol}}
with 
\begin{equation}
\label{eqn_d}
\N(t)=\left(1 -\frac{\delta^2}{\omega^2}(1-\cos 2\omega t)\right)^{-1}.
\end{equation}
For the phase-space dynamics it then follows from (\ref{eqn_Zandz}) and (\ref{comp-sol}) 
\FF{P(t) &=&\N(t)\Big(
P_0\cos \OM t -\frac{Q_0}{\OM}\big(\om+\delta\big)\sin\OM t\Big)\nn\\
Q(t) &=&\N(t)\Big(
Q_0\cos \OM t +\frac{P_0}{\OM}\big(\om-\delta\big)\sin\OM t\Big),\label{PQ-sol}}
for initial conditions $P(0)=P_0$, $Q(0)=Q_0$. Through the time dependence of $d(t)$ in (\ref{eqn_d}) we can thus encounter periodic divergences in the phase-space dynamics if $\frac{\delta^2}{\omega^2}\geq\frac{1}{2}$, that is for $\delta^2\geq\om^2$, while the dynamics stays bounded otherwise. Figure \ref{fig_PQdyn} shows examples of the phase-space dynamics for different initial conditions and parameter values, in the region were the dynamics stays bounded. For increasing $\delta$ the phase-space trajectory, which is an ellipse in the Hermitian limit ($\delta=0$), develops a characteristic dumbbell shape, symmetric with respect to point reflection $(P,Q)\to-(P,Q)$. The enclosed phase-space area grows with increasing $\delta$, and  above the critical value $\delta_{\rm crit}=\om$ the trajectories extend to infinity. 

\begin{figure}[htb]
\begin{center}
\includegraphics[width=0.32\textwidth]{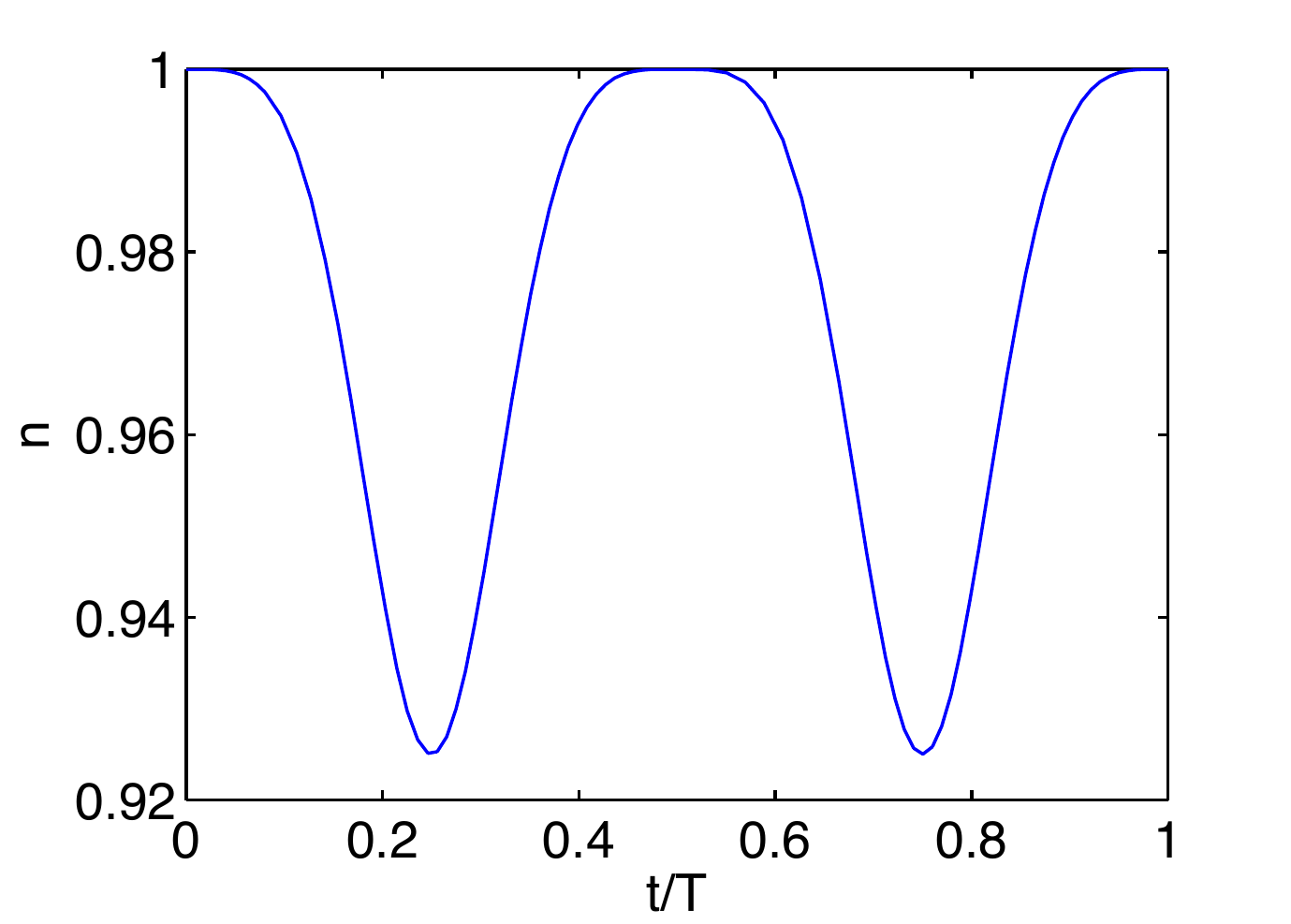}
\includegraphics[width=0.32\textwidth]{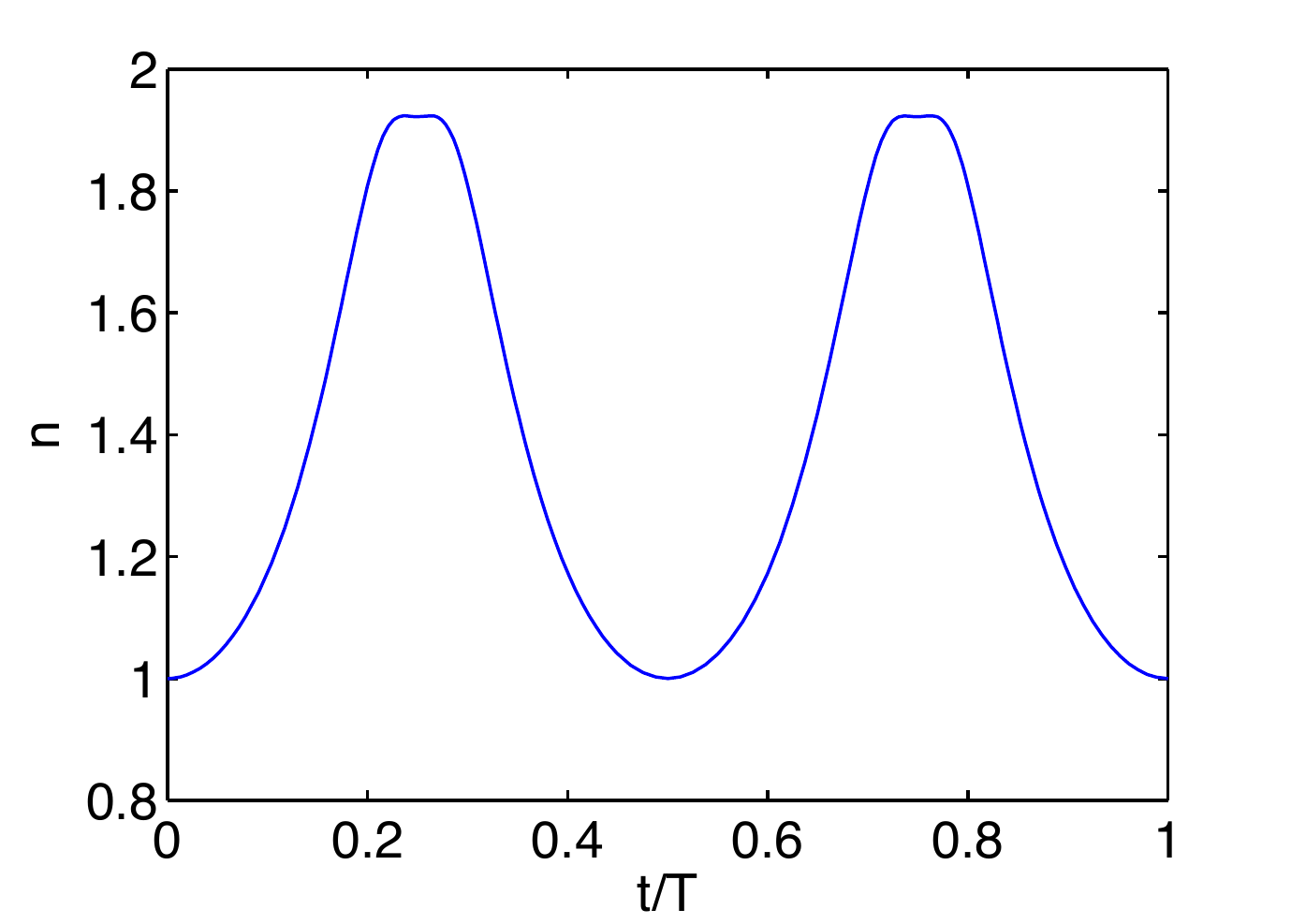}
\includegraphics[width=0.32\textwidth]{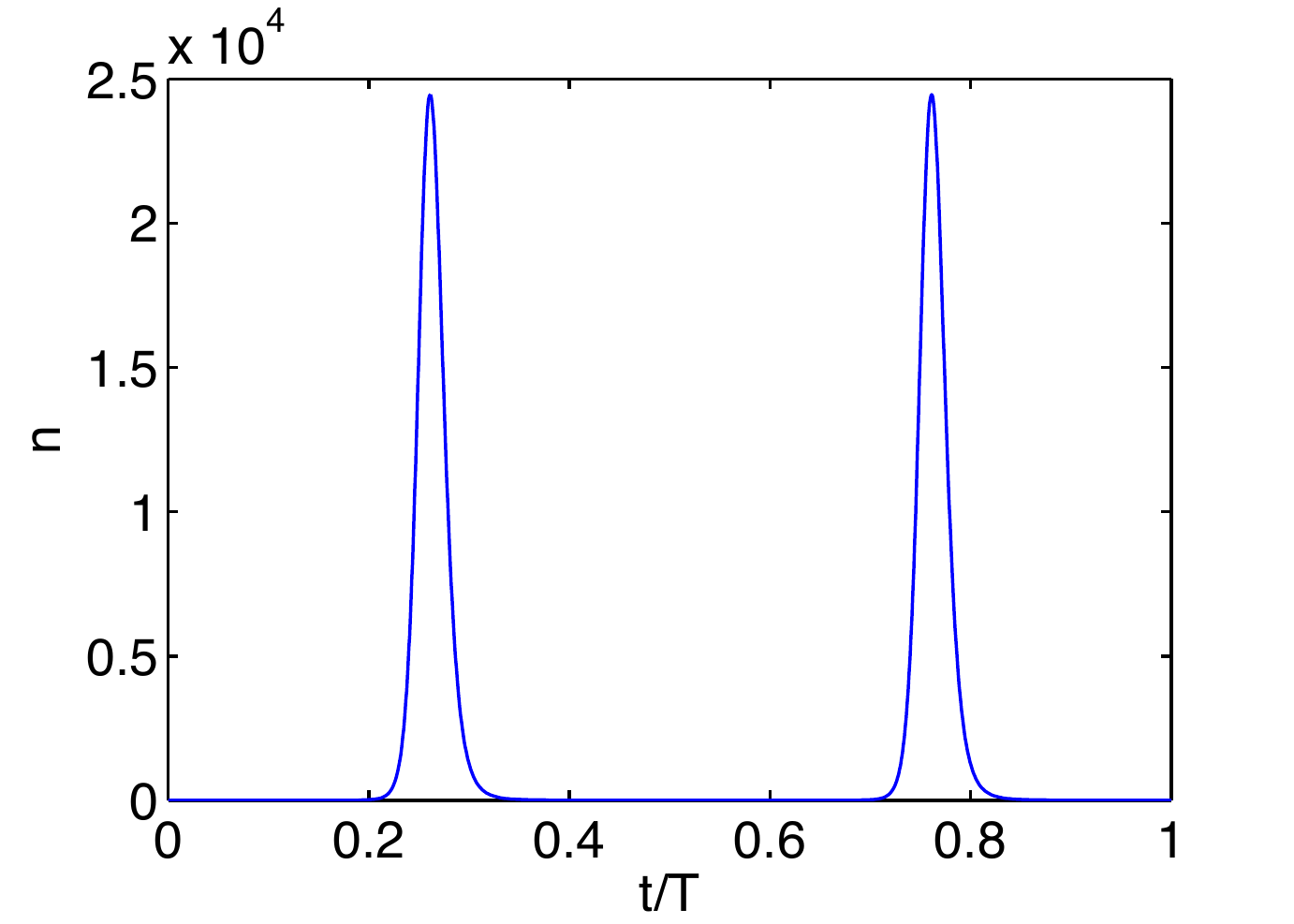}
\caption{Dynamics of the probability $n(t)$ (\ref{dgl-norm}) for different parameters and initial values. We have $P_0=1$ and $\om=1$ for all figures, $Q_0=0$ and $\delta=0.5$ on the left, $Q_0=0$ and $\delta=0.9$ in the middle and $Q_0=1$ and $\delta=0.9$ on the right. Time is in units of the period $T=2\pi/\omega$. }
\label{fig_normdyn}
\end{center}
\end{figure}

The corresponding dynamics of the overall probability as a function of time is given by 
\begin{equation}
n(t)=\sqrt{d(t)}\,\rme^{\,\frac{\delta d(t)}{2\omega^2}[\{(\delta-\om)P_0^2
+(\delta+\om)Q_0^2\}(1-\cos 2\omega t)-2\omega P_0Q_0\sin 2\omega t]}.
\label{n-sol}
\end{equation}
In figure \ref{fig_normdyn} this is plotted over one period for three examples. We observe oscillations that are typical for $PT$-symmetric systems, depending on the initial conditions, the survival probability can decrease or increase periodically. Above the critical value $\delta_{\rm crit}$ it diverges periodically. 

For different initial conditions for the phase-space metric $G$ we observe similar divergences for different parameter values. From the complex phase-space dynamics (\ref{comp-sol}) it is obvious that the origin of the divergences is in the dynamics of the phase-space metric $G$. We can deduce the critical parameter values and initial conditions $G(0)$ for which divergences appear from a geometric interpretation of the dynamics of the metric. 

\subsection{Geometry of the phase-space metric dynamics}
We can parameterise the metric $G$ as 
\begin{equation}
G=\begin{pmatrix}
z-x & y\\
y & z+x
\end{pmatrix},
\end{equation}
with $x,y,z\in\mathds{R}$ and $z$ positive. 
The eigenvalues of $G$ only depend on $z$:
\FF{\lambda_\pm=z\pm \sqrt{z^2-1}\,.\label{G-eig}}
The condition $\det G=1$ then constrains $x,y,z$ to a hyperboloid,
\begin{eqnarray}
z^2-y^2-x^2=1\,.\label{hyper}
\end{eqnarray}
The dynamical equations for the metric in terms of the new variables become 
\begin{eqnarray}
\dot x&=&2 y(\om +\delta x)\nn\\
\dot y&=&-2\om x + 2\delta(1+ y^2)\label{dgf-metric-xyz}\\
\dot z&=&2\delta y z.\nn
\end{eqnarray}
It can be verified by a short calculation that the dynamics (\ref{dgf-metric-xyz}) is confined to planes 
\begin{equation}
\label{z-plane}
z=\frac{z_0}{\om+\delta x_0}\left(\om+\delta x\right).
\end{equation}
in the $(x,y,z)$ space. 
The metric trajectory is thus given by the intersection of a two-sheeted hyperboloid and a plane. If the absolute value of the slope of the plane is larger than one, the resulting curve is a hyperbola, while it is an ellipse otherwise. 
Thus, we find the critical condition
\begin{equation}
\label{eqn_div_zx}
\frac{\delta z_0}{\om+\delta x_0}=\pm1,
\end{equation}
where the plus holds for positive $\delta$, and the minus for negative, that divides the regions in parameter space for which the metric stays bounded from the regions in which one of the eigenvalues of the metric diverges dynamically, while the other tends to zero. For any finite value of $\delta$ there are thus initial conditions for which divergences occur. The dynamics is still strictly periodic, even for the infinite trajectories, and the divergences occur periodically. 

Note that these divergences happen even though the eigenvalues of the quantum Hamiltonian 
are purely real, and the dynamics can be mapped to the dynamics of a harmonic 
oscillator. The origin of this phenomenon lies in the unboundedness of the mapping to 
the Hermitian system, and the related issue that the eigenfunctions of a non-Hermitian operator do not necessarily form a Riesz basis, as has been frequently discussed in the literature 
(see, e.g.~\cite{Baga10,Heil11,Bend12,Sieg12,Baga13,Baga13b,Bend13,Most13,Krej14,Brod14,Henr14,Henr14b}). The fact that the dynamics shows divergences implies that there cannot be a mapping to a Hermitian system via a bounded operator. Further, it is known that a non-trivial pseudo-spectrum as that of the Swanson oscillator means that not only is there no bounded mapping to an isospectral Hermitian system, but also the time evolution operator cannot be bounded \cite{Krej14}. Whether the lack of an unbounded mapping to a Hermitian counterpart, on the other hand, always implies the occurrence of divergences in the classical dynamics is a non-trivial question that goes beyond the scope of the present paper. 
\section{Quantum dynamics for Gaussian wave packets} 
\label{sec_Qdyn}
From the classical dynamics discussed previously one can directly deduce the quantum dynamics of Gaussian wave packets in a semiclassical approximation. As in Hermitian quantum mechanics, this approximation becomes exact if the Hamiltonian is of no higher order than quadratic in position and momentum. Thus, the previous results allow us to deduce the exact quantum dynamics for initial Gaussian states. See \ref{appendix_Gaussian_dyn} for further details. 

Consider a  family of Gaussian states
\begin{equation}\label{eq:coh_state_r}
\psi_Z^B(x)= \left(\frac{\Im B}{\pi}\right)^{1/4} {\rm e}^{{\rm i}[P(x-Q)+\frac{1}{2} B (x-Q)^2]} 
\,\, ,
\end{equation}
with $Z=(P,Q)\in \mathds{R}\times \mathds{R}$, and  $B\in\mathds{C}$ with positive imaginary part, $\Im B>0$. This last  condition ensures that the state  is in $L^2(\R)$ and the prefactor is chosen such that the state is normalised to one. The uncertainty parameter $B$ is related to the covariance matrix
\begin{equation}
\label{eq:defG}
G=\frac{1}{\Im(B)}\begin{pmatrix}1&-\Re(B)\\
-\Re(B)&\Re(B)^2+\Im(B)^2
\end{pmatrix},
\end{equation}
such that the expectation values 
$\langle \hat A\rangle=\langle \psi|\hat A|\psi\rangle/\langle \psi|\psi\rangle$
of an observable $\hat A$ and its variance are
given by
\begin{eqnarray}
\label{nh-class-A}
\langle \hat A\rangle\approx A(Z) \ ,\quad
(\Delta \hat A)^2\approx\tfrac12 \nabla A\cdot G^{-1}\nabla A,
\end{eqnarray}
up to orders of $\hbar$ or $\hbar^2$, respectively, where $A(Z)$ is
the Weyl representation of $\hat A$. In particular we have
$(\Delta \hat p)^2=g_{qq}/2$ and $(\Delta \hat q)^2=g_{pp}/2$, i.e.~the
diagonal elements of the metric $G$ represent the uncertainties of
momentum and position. 
The eigenvalues 
\FF{g_\pm=\tfrac12 \Big(g_{pp}+g_{qq} \pm \sqrt{(g_{pp}+g_{qq})^2-4}\,\Big)\label{G-eig-g}}
of the metric  are equal to the uncertainties in a frame rotated by
an angle 
\FF{\varphi=\tfrac12 \arctan \frac{2g_{pq}}{g_{pp}-g_{qq}}\,.\label{G-rot}}
Note that we have $G=I$ for a standard Glauber coherent state.

\begin{figure}[htb]
\begin{center}
\includegraphics[width=0.32\textwidth]{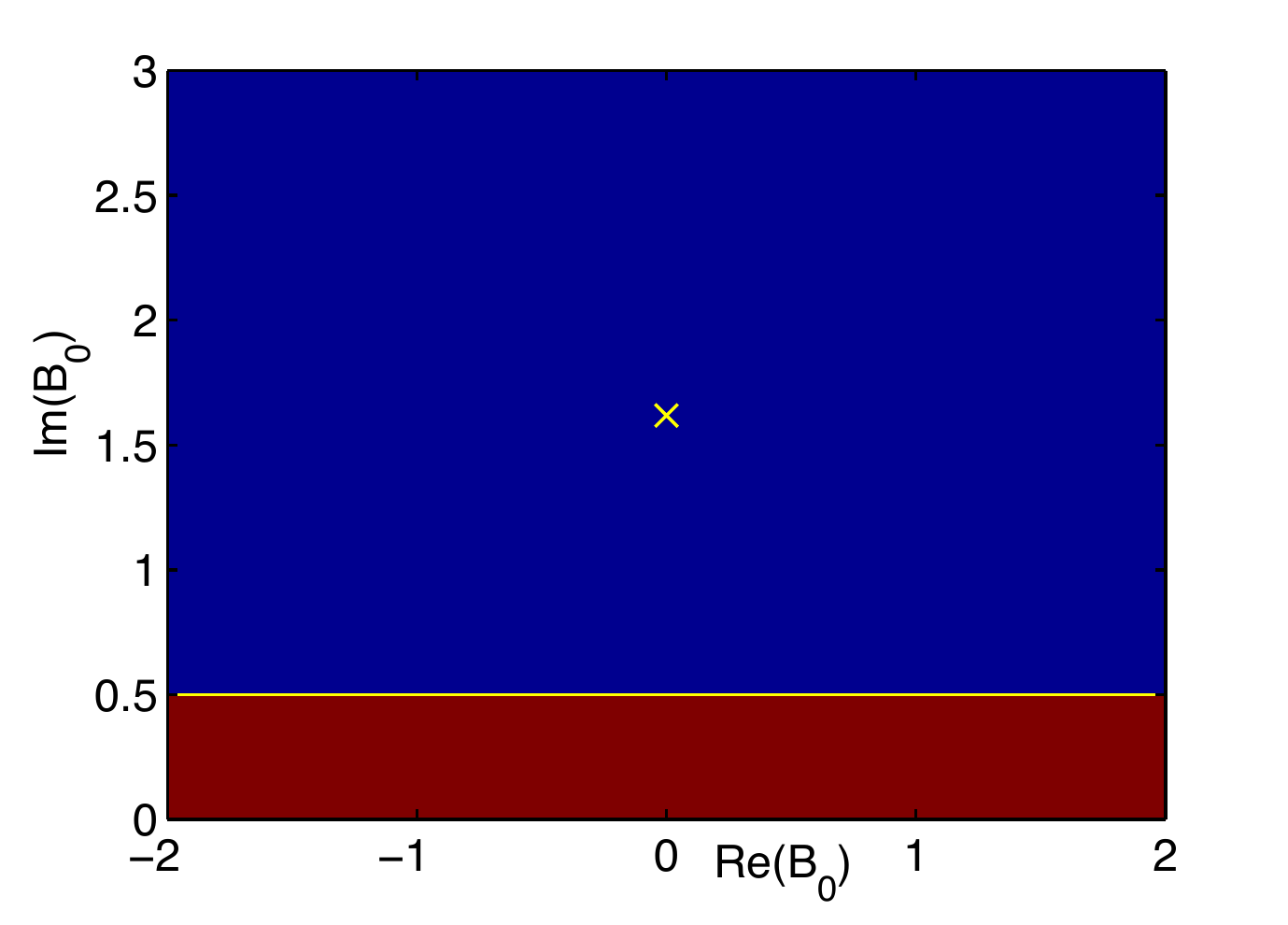}
\includegraphics[width=0.32\textwidth]{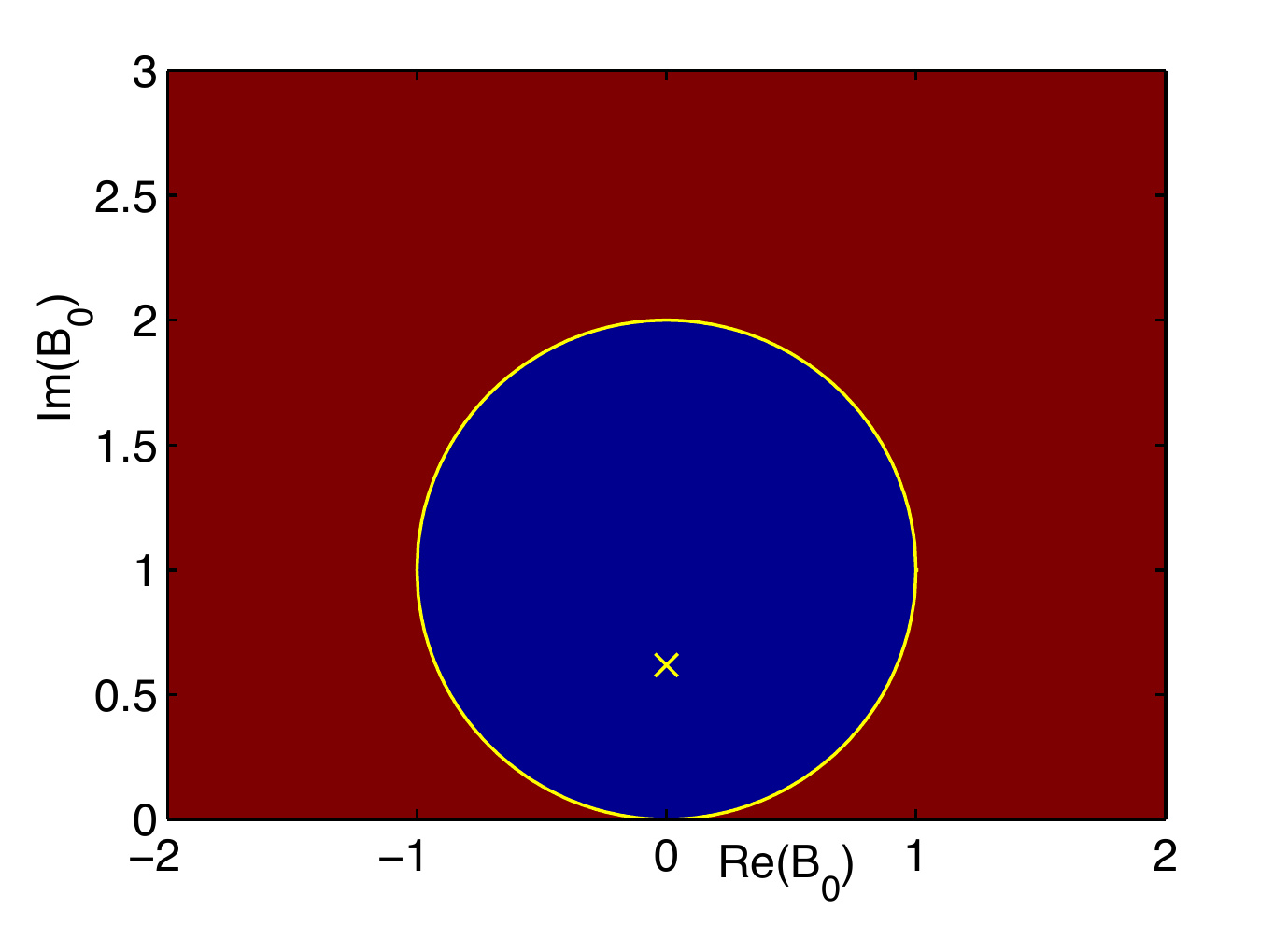}
\includegraphics[width=0.32\textwidth]{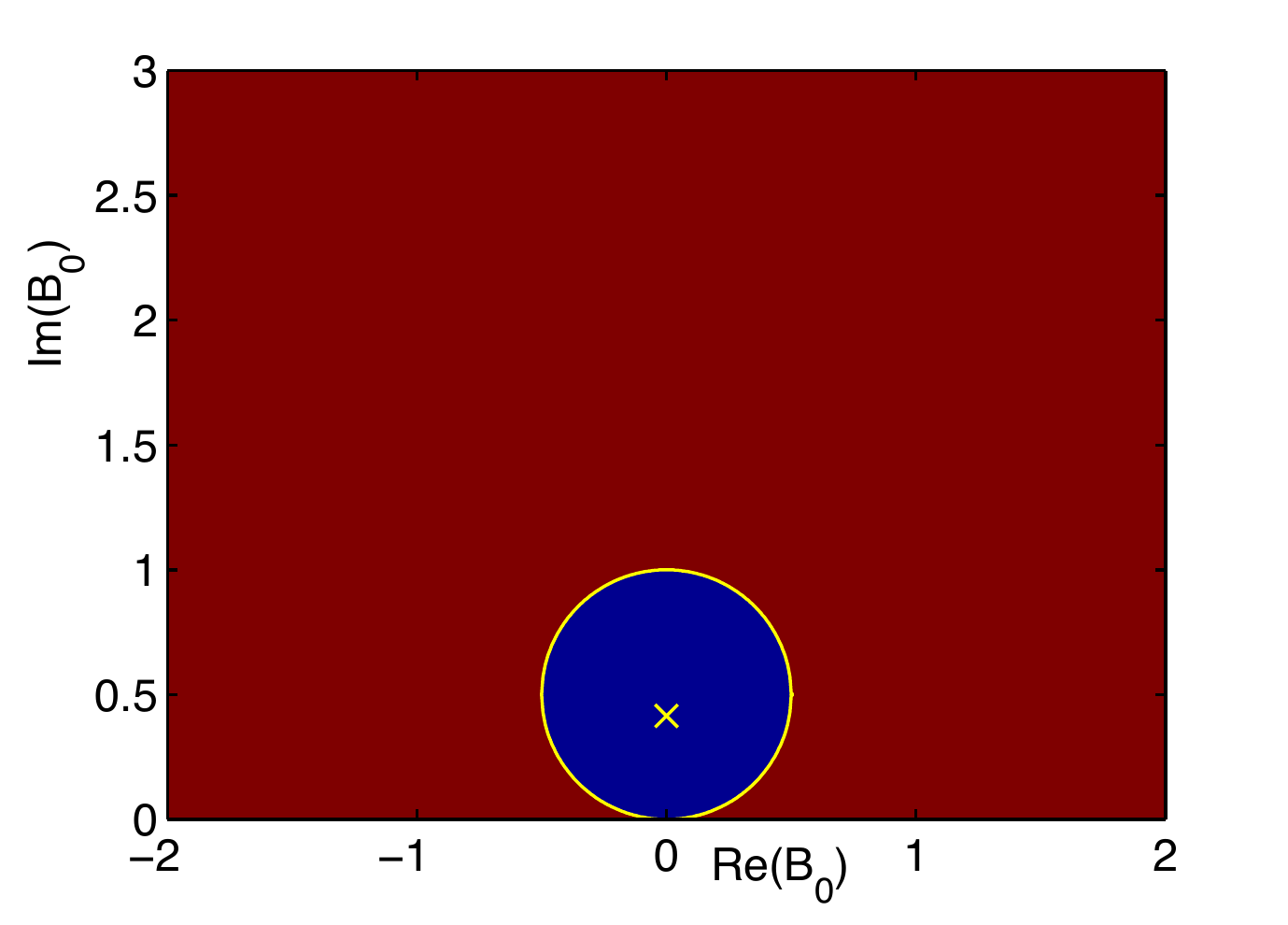}
\caption{Regions of the complex uncertainty parameter $B$ for which initial wave packets diverge (red) or stay Gaussian (blue) for parameter values $\om=1$ and $\delta=0.5$ (left), $\delta=-0.5$ (centre), and $\delta=-1$ (right). The yellow cross marks the normalisable ground state.}
\label{fig_div_region}
\end{center}
\end{figure}

An initial Gaussian wave packet (\ref{eq:coh_state_r}) stays dynamically Gaussian with time-dependent parameters $Z(t)$ and $B(t)$ under the time evolution generated by a quadratic Hamiltonian. The time dependence of the centre $Z=(P,Q)$ and the parameter $B$ follow the classical dynamics discussed in the previous section. The ground state of the system appears as a fixed point of this classical dynamics. We can now see that the divergence of one of the eigenvalues of $G$ means that the quantum mechanical wave packet spreads infinitely and is momentarily no longer in $L^2(\R)$. For any parameter values we can find initially Gaussian wave packets that dynamically diverge in this way. The initial value of $B$ for which this happens can be directly deduced from condition (\ref{eqn_div_zx}), using that
\begin{eqnarray}
x=\frac{1}{2}\left(g_{qq}-g_{pp}\right)&=&\frac{1}{2\Im(B)}\left(\Re(B)^2+\Im(B)^2-1\right)\\
z=\frac{1}{2}\left(g_{qq}+g_{pp}\right)&=&\frac{1}{2\Im(B)}\left(\Re(B)^2+\Im(B)^2+1\right).
\end{eqnarray}
From this we find that for positive $\delta$  initially Gaussian wave packets with $\Im(B_0)<\frac{\delta}{\omega_0}$ diverge in finite time, whereas the others stay convergent. For negative $\delta$ Gaussian wave packets whose initial value of $B$ lies in the circle in the complex plane defined by 
\begin{equation}
\Re(B_0)^2+\left(\Im(B_0)+\frac{\om}{2\delta}\right)^2=\frac{\om^2}{4\delta^2}\label{eqn_circle_B}
\end{equation}
stay in $L^2$ for all times, while those outside diverge in finite time. In the Hermitian limit $\delta\to0$ all initial Gaussian wave packets dynamically stay in $L^2(\mathds{R})$ as expected. For every finite value of $\delta$, however, there are initial Gaussian wave packets that diverge in finite time. Examples of the regions of the initial parameter $B$ in the upper halfplane for which initial Gaussian wave packets diverge are shown in figure \ref{fig_div_region}. Note that the relation between the conditions for negative and positive $\delta$ can be understood in the following way. The quantum Hamiltonian (\ref{eqn_Ham}) with positive $\delta$ is related to that with negative $\delta$ by conjugation with a Fourier transform $\hat U=\rme^{-\frac{\rmi}{4}\left(\hat p^2+\hat q^2\right)}$, that maps $\hat p\to-\hat q$, and $\hat q\to\hat p$. The action of this unitary transform onto the $B$ parameter of a wave packet according to equation (\ref{tildeB_app}) is given by $\tilde B=\Omega_* B:=-\frac{1}{B}$. This transformation maps the curve $\Im(B_0)+\rmi \frac{\delta}{\om}$, which forms the boundary between dynamically diverging and non-diverging wave packets for positive $\delta$ to the curve (\ref{eqn_circle_B}) for negative $\delta$.

The dynamics of an initial wave packet (\ref{eq:coh_state_r}) can also be obtained directly from the approach of \ref{appendix_Gaussian_dyn} using the equations  
\begin{eqnarray}
Z(t)&=\Re(S(t)Z_0)-\Omega G(t)\Im(S(t)Z_0)\\
B(t)&=S(t)_*B_0:=\frac{S(t)_{pp}B_0+S(t)_{pq}}{S(t)_{qp}B_0+S(t)_{qq}},
\end{eqnarray}
where $G(t)$ is defined via $B(t)$ according to equation (\ref{eq:defG}), and $S(t)\in Sp(n,\C)$ is the solution of 
\begin{equation}
\dot S=\Omega (H''-\rmi\Gamma'') S\,\, ,\quad \text{with}\quad S(0)=I.
\end{equation}
In this formulation the time dependent uncertainty parameter $B(t)$ appears as a M\"obius transformation of the initial $B_0$ with time dependent parameters.

Let us finally comment on the role of the unbounded mapping (\ref{eqn_eta}) to the Hermitian oscillator (\ref{eqn_HO}), which can also be used to obtain the quantum dynamics. The time dependent state of the Swanson Hamiltonian is given by 
\begin{equation}
\label{eqn-mapped-dyn}
\psi(t)=\hat\eta\,\rme^{-\rmi \hat H_{herm}t}\hat\eta^{-1}\psi(0),
\end{equation}
where the unboundedness of $\hat\eta$ (and also $\hat\eta^{-1}$) can lead to divergences of an initially normalisable state. It is further obvious from (\ref{eqn-mapped-dyn}) that the dynamics is periodic with the period of the Hermitian harmonic oscillator. For the case of an initial Gaussian coherent state (\ref{eqn-mapped-dyn}) can be explicitly calculated using the approach summarised in \ref{appendix_Gaussian_dyn}, which provides an alternative way of reproducing the presented quantum dynamics, and in particular equations (\ref{G-sol}), (\ref{PQ-sol}). To deduce the time dependent norm in equation (\ref{n-sol}) it is more convenient to use the Wigner function as in \ref{app-semi} as this avoids the unecessary calculation of additional phase factors. 
\section{Conclusion and perspectives}
We have analysed a classical version of the non-Hermitian Swanson
oscillator and derived closed form results for the phase-space
metric and the phase-space dynamics, which were found to be periodic
in time. The results exactly capture the dynamics of 
quantum expectation values for initial Gaussian wave packets. 
We observed that the classical metric and trajectories and the
quantum wave functions diverge in a finite time if the parameters are chosen in a critical region, even though the eigenvalues of the quantum system are purely real. 
This is, of course, not true for all initial states. All quantum
eigenstates of the Hamiltonian and their finite linear combinations for example
will stay bounded.  We have concluded that the divergences are related to the unboundedness of the operator mapping the system to an isospectral Hermitian system. Whether the lack of an unbounded mapping to a Hermitian counterpart on the other hand generically leads to divergences of Gaussian wave packets and the corresponding classical dynamics is an important question for future investigations. 

Let us conclude with a conjecture concerning the physical origin of the divergences. It should be noted that the Swanson Hamiltonian considered here has a positive imaginary part that does not vanish for $|p|,|q|\to\infty$. Physically this corresponds to a situation where an infinite amount of probability is fed into the system. It is very suggestive that this is the origin of the observed divergences. It would be natural to expect that Hamiltonians whose imaginary parts are integrable functions on phase space will not lead to similar divergences, and, in the presence of  $PT$-symmetry, a bounded mapping to a Hermitian counterpart would be expected. It has been proven in \cite{Addu12,Addu12b,Mity13}, for example, that certain well-behaved types of non-Hermitian perturbations of Hermitian operators with ``harmonic oscillator type'' spectra will lead to Riesz bases, an no finite time divergences would be expected. The systematic investigation of the connection between such mathematical constraints and their physical interpretation would be an interesting topic for further studies.

\section*{Acknowledgments}
EMG gratefully acknowledges support via the Imperial College JRF scheme and the L'Or\'eal UNESCO Women in Science programme. AR acknowledges support from an EPSRC DTA grant. The research was supported in part by EPSRC Mathematics Platform grant EP/I019111/1. The authors would like to thank Hugh Jones for stimulating discussions. 
\appendix
\section{Time evolution of Gaussian states generated by (complex) quadratic Hamiltonians}
\label{appendix_Gaussian_dyn}
Here we give a brief summary of the results obtained in \cite{12complex_coherent} on the exact time evolution of Gaussian wave packets generated by complex quadratic Hamiltonians. For convenience we restrict the discussion to the one-dimensional case of interest here.  

We are interested in the solutions of the Schr\"odinger equation with a Hamiltonian that is maximally quadratic in $\hat p$ and $\hat q$ with complex coefficients and an initially Gaussian state. The quantum Hamiltonian can be formulated as $\hat{\mathcal{H}}=\mathcal{H}(\hat p,\hat q)$, where $\mathcal{H}(p,q)$ is a classical Hamiltonian function, which is a polynomial of order two in $p$ and $q$, where we assume the symmetrised version $pq\leftrightarrow \frac{1}{2}(\hat p\hat q+\hat q \hat p)$. It can be verified by direct calculation that during the time evolution an initially coherent state stays of the form 
\begin{equation}\label{eq:coh_state_appb}
\psi_z^B(x)=\left(\frac{\Im B}{\pi}\right)^{1/4} {\rm e}^{{\rm i}\gamma(t)} {\rm e}^{{\rm i}[p(x-q)+\frac{1}{2} B (x-q)^2]} 
\,\, ,
\end{equation}
with $z=(p,q)\in \mathds{C}\times \mathds{C}$, $\gamma\in\mathds{C}$, and $B\in\mathds{C}$, by inserting this ansatz with time dependent parameters into the Schr\"odinger equation. Note that only if $\Im B>0$ this state is normalisable and is Gaussian in the conventional sense. Note further that two wave functions of the form (\ref{eq:coh_state_appb}) with different complex values of $p$ and $q$ can physically correspond to the same state, i.e., differ only by a phase \cite{12complex_coherent}. 

Inserting the ansatz (\ref{eq:coh_state_appb}) into the Schr\"odinger equation leads to the dynamical equations for the parameters
\begin{eqnarray}
\label{dyn_appa_B}
\dot{B}=-\mathcal{H}_{pp}B^2-2\mathcal{H}_{pq}B-\mathcal{H}_{qq}\\
B\mathcal{H}_p+\mathcal{H}_q=B\dot{q}-\dot{p}\label{dyn_appa_pq}\\
\dot{\gamma}=p\dot{q}+\frac{\rmi}{4}\left(\mathcal{H}_{pp}B-\mathcal{H}_{qq}B^{-1}\right)-\mathcal{H}\label{dyn_appa_gamma},
\end{eqnarray}
where $\mathcal{H}_q$ denotes the derivative of the corresponding classical Hamiltonian function with respect to $q$ etc, and we have used that $\mathcal{H}_{pq}=\mathcal{H}_{qp}$. 
The second derivatives $\mathcal{H}_{qp},\ \mathcal{H}_{qq},\ \mathcal{H}_{pp}$ are constant and thus equation (\ref{dyn_appa_B}) for the uncertainty parameter $B$ is a standard Riccati equation that can be solved independently of the remaining equations. Equation (\ref{dyn_appa_pq}) encodes the dynamics of the phase space coordinates $p$ and $q$. Equation (\ref{dyn_appa_gamma}) describes the time evolution of the complex phase (incorporating the norm) of the wave packet and can be integrated once  the first two equations have been solved.

There are an infinite number of dynamical equations for $q$ and $p$ that are compatible with equation (\ref{dyn_appa_pq}), relating to the ambiguity of states with complex centres $p,q$. Two natural choices are of particular relevance here; the complexified version of Hamilton's canonical equations, and the dynamical equation arising from the assumption that $p$ and $q$ are real. Let us start with the first choice, Hamilton's equations of motion,
\begin{equation}
\dot q=\frac{\partial \mathcal{H}}{\partial p},\quad \dot p=-\frac{\partial\mathcal{H}}{\partial q},
\end{equation} 
for which $p(t)$ and $q(t)$ are in general complex. 
Similar to the real case, it is straight forward to verify that this set of equations is solved by
\begin{equation}
\label{eqn_zoft_app}
z(t)=Sz(0),
\end{equation}
where $z=(p,q)^T$ denotes the complex phase space vector and the symplectic matrix 
\begin{equation}
\label{Sdot_app}
S=\begin{pmatrix}S_{pp}& S_{pq}\\
S_{qp}& S_{qq}\end{pmatrix},\quad {\rm with}\quad S^{\rm T}\Omega S=\Omega,\quad {\rm where}\quad \Omega=\begin{pmatrix}0&-1\\1&0\end{pmatrix},
\end{equation} 
is defined as the solution of the dynamical equation
\begin{equation}
\dot S=\Omega \mathcal{H}'' S, 
\end{equation}
where $S(0)$ is the identity, and $\mathcal{H}''$ denotes the Hessian matrix of second phase space derivatives of the complex Hamiltonian function. The matrix $S$ also directly yields the time dependence of $B$ via
\begin{equation}
\label{eqn_Boft_app}
B=S_*B(0):=\frac{S_{pp}B(0)+S_{pq}}{S_{qp}B(0)+S_{qq}},
\end{equation}
which is also straight forwardly verified by a simple calculation. It is interesting that the time evolution of $B$ in (\ref{eqn_Boft_app}) has the form of a M\"obius transformation. Note that the time evolution presented here is a direct complex generalisation of 
Gaussian wave packet dynamics generated by real Hamiltonians, as discussed in detail in \cite{Litt86}. 

Equations (\ref{eqn_zoft_app}) and (\ref{eqn_Boft_app}) together with the solution of (\ref{dyn_appa_gamma}) parameterise the solution of the time dependent Schr\"odinger equation, that is, the time dependent wave function. If one is interested in the (real) expectation values of position and momentum, however, these have yet to be extracted from the complex phase space variables and the uncertainty parameter $B$. 
Explicitly evaluating the expectation values in a wave function of the form (\ref{eq:coh_state_appb}) yields
\begin{eqnarray}
\label{eqn_Pandp_app}
P:=\langle \hat p\rangle=\Re(p)-\frac{\Re(B)}{\Im(B)}\Im(p)+\frac{\Re(B)^2+\Im(B)^2}{\Im(B)}\Im(q)\\
Q:=\langle \hat q\rangle=\Re(q)-\frac{1}{\Im(B)}\Im(p)+\frac{\Re(B)}{\Im(B)}\Im(q)\,.
\end{eqnarray}
With the definition $Z=(P,Q)^T$,
\begin{equation}
\label{eqn-GofB}
G=\frac{1}{\Im(B)}\begin{pmatrix}1&-\Re(B)\\
-\Re(B)&\Re(B)^2+\Im(B)^2
\end{pmatrix},
\end{equation}
and the standard symplectic matrix $\Omega$ as in (\ref{Sdot_app}), 
these equations take the compact form 
\begin{equation}
Z=\Re\,z-\Omega G\,\Im\,z.
\end{equation}

On the other hand, assuming $p$ and $q$ to be real in (\ref{dyn_appa_pq}) directly leads to equations of motion for $P$ and $Q$ given by
\begin{equation}
\dot Z=\Omega\nabla \Re(\mathcal{H}) +G^{-1}\nabla \Im(\mathcal{H}),
\end{equation}
which are dynamically coupled to the time evolution of the metric $G$ that follows from the dynamical equation for $B$ as
\begin{equation}
\label{eqn-dotG_app}
\dot G=G\Omega^T\Im(\mathcal{H})\Omega G-G\Omega \Re(\mathcal{H})''+\Re(\mathcal{H})''\Omega G-\Im(\mathcal{H})''.
\end{equation} 
We can solve equation (\ref{eqn-dotG_app}) in a similar way to equation (\ref{dyn_appa_B}) as
\begin{equation}
G=\Phi_*G(0),
\end{equation}
where $\Phi$ solves the equation
\begin{equation}
\dot\Phi=\Omega_4 K \Phi,
\end{equation}
where $\Omega_4$ is the symplectic matrix on a four dimensional phase space and we have defined the block matrix
\begin{equation}
K=\begin{pmatrix}-\Omega^T\Im(\mathcal{H})''\Omega & \Omega\Re(\mathcal{H})''\\
-\Re(\mathcal{H})''\Omega  & \Im(\mathcal{H})''
\end{pmatrix},
\end{equation}
and where $\Phi(0)$ is the identity.

The action of an operator 
\begin{equation}
\label{eqn_eta_app}
\hat \eta= \rme^{-\frac{\rmi}{2}\left(M_{pp}\hat p^2+M_{qq}\hat q^2+M_{pq}\left(\hat p\hat q+\hat q \hat p\right)\right),}
\end{equation}
with $M_{pp},\,M_{qq},\,M_{pq}\in\mathds{C}$ on a Gaussian state, 
can be easily deduced by interpreting 
$\hat \eta$ as a time evolution operator generated by a complex Hamiltonian from time
zero to time $t=1$, and applying the previous results. The state $\hat\eta\psi(x)$, with $\psi(x)$ in (\ref{eq:coh_state_appb}) is again of the form (\ref{eq:coh_state_appb}), with centre $\tilde Z$ and uncertainty parameter $\tilde B$ given by
\begin{eqnarray}
\tilde Z&=\Re(SZ)-\Omega \tilde G\Im(SZ)\\
\tilde B&=S_*B:=\frac{S_{pp}B+S_{pq}}{S_{qp}B+S_{qq}},\label{tildeB_app}
\end{eqnarray}
where the metric $G$ is given in terms of $\tilde B$ according to equation (\ref{eqn-GofB}), and $S$ is the solution of the dynamical equation
\begin{equation}
\label{eqn_dynS}
\dot{\tilde S}=\Omega M \tilde S\,\, ,\quad \text{with}\quad M=\begin{pmatrix}M_{pp}&M_{pq}\\M_{pq}&M_{pp}\end{pmatrix}\quad\text{and}\quad\tilde S(0)=I, 
\end{equation} 
at time $t=1$, i.e.,  $S=\tilde S(1)$. 

We can directly apply this to find the ground state of the non-Hermitian Hamiltonian (\ref{eqn_Ham}) from the ground state of the harmonic oscillator (\ref{eqn_HO}) (i.e. $B=\rmi$ and $(P,Q)=(0,0)$) acted upon by $\hat\eta=\rme^{-\frac{\theta}{2}\left(\hat{p}^{2}-\hat{q}^{2}\right)}$, with $\theta=-\frac{1}{2}{\rm atan}\frac{\delta}{\om}$. 
Equation (\ref{eqn_dynS}) directly integrates to
\begin{equation}
S=\begin{pmatrix}\cos\theta& -\rmi\sin\theta\\
-\rmi\sin\theta& \cos\theta\end{pmatrix},
\end{equation}
with $\theta$ as above.
Thus we find 
\begin{equation}
\tilde B=\rmi\frac{\om}{\omega-\delta}
\end{equation}
and $\tilde Z=(0,0)^T$. 
\section{Semiclassical limit of the quantum dynamics generated by non-Hermitian Hamiltonians}
\label{app-semi}
Here we briefly summarise the derivation of the classical dynamics related to the quantum dynamics generated by non-Hermitian Hamiltonians of \cite{11nhcs}. An elegant way of deriving Hamiltonian dynamics from quantum dynamics in the semiclassical limit is by considering quantum phase space dynamics of initially well localised states, such as Gaussian states. Following the same approach for non-Hermitian Hamiltonians yields the classical phase space dynamics discussed in the present paper for the case of the Swanson Hamiltonian. 

We consider a non-Hermitian Hamiltonian decomposed into its Hermitian and anti-Hermitian parts according to $\hat{\mathcal{H}}=\hat H-\rmi\hat \Gamma$, where $\hat H$ and $\hat \Gamma$ are Hermitian and given by Weyl quantisations of sufficiently well-behaved classical phase space functions. The dynamics of the Wigner function $W$ for an arbitrary state, that follows from the Schr\"odinger equation as 
\begin{equation}
\label{eqn:wigner-evo}
\rmi\hbar\frac{\partial W}{\partial t}=(H\sharp W-W\sharp H)-\rmi(\Gamma\sharp W+W\sharp\Gamma)\,\, , 
\end{equation}  
where $A\sharp B$ denotes the Weyl product for two phase space functions $A(z)$ and $B(z)$:
\begin{equation}
\nonumber
A\sharp B= \sum_{k=0}^{\infty} \frac{1}{k!}\bigg(\frac{\rmi\hbar }{2}\bigg)^k A\big(\overleftarrow{\nabla}_z\cdot \Omega \overrightarrow{\nabla}_{z}\big)^kB\,\, ,
 \end{equation}
 where the arrows over the differential operators indicate whether they act to the left or the right.
 
We are interested in the semiclassical limit of the phase space dynamics, and thus the leading orders in $\hbar$ of (\ref{eqn:wigner-evo}), which yields
\begin{equation}\label{eq:pphase-space-evo}
\hbar \frac{\partial W}{\partial t}=-\bigg(-\frac{\hbar^2}{4}\Delta_{\Gamma}- \hbar \nabla H\cdot\Omega \nabla+2\Gamma \bigg)W\, ,
\end{equation}
where we have defined $\Delta_{\Gamma}=\nabla\cdot \Omega^T\Gamma''\Omega\nabla$. 
This equation is the non-Hermitian equivalent of the Liouville equation for Hamiltonian systems. To deduce from this a phase space dynamics of individual phase space points, that generalises Hamilton's canonical equation of motion, we need to consider the dynamics of well localised states, that can be identified with phase space points in the semiclassical limit. An ideal choice are Gaussian states of the form (\ref{eq:coh_state_appb}). The Wigner function of such a Gaussian state is itself Gaussian, of the form 
\begin{equation}
\label{eqn:wigner_gauss}
W(\zeta)=\frac{1}{\pi\hbar}\rme^{-\frac{1}{\hbar} (\zeta-Z)\cdot G(\zeta-Z)}.
\end{equation}
where the co-variance matrix $G$ is given by (\ref{eqn-GofB}) and where $\zeta$ denotes the phase space vector. This distribution does indeed shrink to a phase space point $Z$ in the limit $\hbar\to0$, where $G$ appears as a local metric. 

Following the approach of \cite{Hell75,Litt86} for Hermitian systems to derive the classical dynamics, 
we insert a Gaussian ansatz (with a time-dependent norm)
\begin{equation}
W(\zeta,t)=\frac{n(t)}{\pi\hbar}\rme^{-\frac{1}{\hbar} (\zeta-Z(t))\cdot G(t)(\zeta-Z(t))}
\end{equation}
for the time-dependent Wigner function into the dynamical equation (\ref{eq:pphase-space-evo}) and Taylor expand the Hamiltonian around its centre up to second order. This yields the following dynamical equations 
\begin{eqnarray}
\dot Z&=\Omega\nabla H(Z)-G^{-1}\nabla\Gamma(Z)\label{eq:theEq1}\\
\dot G&=H''(Z)\Omega G-G\Omega H''(Z)+\Gamma''(Z)-G\Omega^T\Gamma''(Z)\Omega G\label{eq:theEq2}\\ 
\dot n &=-(\,\frac{2}{\hbar}\Gamma(Z)+\frac{1}{2}\tr[\Omega^T\Gamma''(Z)\Omega G]\,)\, n\,\, .
\end{eqnarray}
for the centre $Z$, the metric $G$, and the norm $n$ of the quantum state.
%%%
%
%

\end{document}